\documentclass[sigconf,screen]{acmart}
\usepackage[utf8]{inputenc}
\usepackage{tikz}
\usepackage{amsmath}
\usepackage{filecontents}
\usepackage[flushleft]{threeparttable}
\usepackage{amsmath,amsfonts}
\usepackage{graphicx}
\usepackage{textcomp}
\usepackage{url}
 
\usepackage{lipsum,tabularx}
\usepackage{pifont}
\usepackage{multicol}
\usepackage{multirow}
\usepackage{colortbl}
\usepackage{booktabs}
\usepackage{setspace}

\hypersetup{
  linkcolor={blue!70!black},
  citecolor={red!70!black},
  urlcolor={blue!70!black}
}
\def\Snospace~{\S{}}

\usepackage[T1]{fontenc}

\usepackage{algorithm}
\usepackage{algorithmicx}
\usepackage[noend]{algpseudocode}
\usepackage{balance}

\usepackage{bm}
\usepackage{fp}
\usepackage{siunitx}
\usepackage{amsthm}
\usepackage[labelfont=bf,font=small,skip=5pt]{caption}
\usepackage{subcaption}

\usepackage{comment}

\usepackage{fancyhdr}
\usepackage{tikz}

\usepackage{xspace}

\newcommand{\boxbeg}{
  \vspace{2px}
  \noindent\begin{tabular}{|l|}\hline
    \begin{minipage}{3.2in}
      \vspace{2px}
      \noindent
      }

      \newcommand{\boxend}{
      \vspace{2px}
    \end{minipage} \\ \hline
  \end{tabular}
  \vspace{-10pt}
}

\usepackage{algpseudocode}
\usepackage[htt]{hyphenat}

\usepackage{wrapfig}

\usepackage[scaled=.7]{beramono}

\usepackage{subcaption}

\usepackage{enumitem}

\usepackage{booktabs}
\usepackage{multirow}
\usepackage{makecell}
\usepackage{tabularx}
\usepackage{enumitem}
\usepackage{placeins}
\usepackage{amsmath}
\usepackage{subcaption}
\usepackage{microtype}
\usepackage{xcolor}
\usepackage{adjustbox}
\usepackage{cuted}
\usepackage{colortbl}
\usepackage{dblfloatfix} 
\usepackage{float}
\usepackage{caption}
\usepackage{array}

\definecolor{XLSXBlue}{HTML}{002060}
\definecolor{XLSXRed}{HTML}{C00000}
\definecolor{XLSXGray}{HTML}{F2F2F2}

\usepackage{enumitem}
\usepackage{mdframed}
\newcounter{finding}
\newcommand{\finding}[1]{\refstepcounter{finding}
    \vspace{0.5mm}
    \begin{mdframed}[linecolor=gray,roundcorner=12pt,backgroundcolor=gray!15,linewidth=3pt,innerleftmargin=2pt, leftmargin=0cm,rightmargin=0cm,topline=false,bottomline=false,rightline = false]
    \textbf{Finding \arabic{finding}:} #1
    \end{mdframed}
    \vspace{0.5mm}
}
\newcommand{\upd}[2]{#2}
\newcommand{\updno}[2]{#2}
\AtBeginDocument{%
  }

\copyrightyear{2026}
\acmYear{2026}
\setcopyright{cc}
\setcctype{by}
\acmConference[ASE '26]{Proceedings of the 41st IEEE/ACM International Conference on Automated Software Engineering}{October 12--16, 2026}{Munich, Germany}
\acmBooktitle{Proceedings of the 41st IEEE/ACM International Conference on Automated Software Engineering (ASE '26), October 12--16, 2026, Munich, Germany}
\acmDOI{10.1145/3832783.3837456}
\acmISBN{979-8-4007-2882-2/2026/10}

\begin{document}

\newcommand{\sys}{\textsc{AIRFLOW}\xspace} %CREATE %TSER
% \title{Semantics Stay but Fairness Shifts: Format Effects in LLM Document Workflows}
\title{Not as Sweet by Another Name: An Empirical Study of Format Robustness in LLM Document Workflows}
% Testing DL library APIs via context similarity

\author{Xiaoyu Zhang}
\affiliation{%
  \institution{Nanyang Technological University}
  \city{Singapore}
  \country{Singapore}
}
\email{xiaoyu.zhang@ntu.edu.sg}

\author{Xianyun Cheng}
\affiliation{%
  \institution{National University of Singapore}
  \city{Singapore}
  \country{Singapore}
}
\email{xianyuncheng0909@gmail.com}

\author{Tianlin Li}
\authornote{Corresponding author.}
\affiliation{%
  \institution{Beihang University}
  \department{State Key Laboratory of Complex \& Critical Software Environment}
  \city{Beijing}
  \country{China}
}
\email{tianlin001@buaa.edu.cn}

\author{Yuwei Zheng}
\affiliation{%
  \institution{Beihang University}
  \department{State Key Laboratory of Complex \& Critical Software Environment}
  %  (SKLCCSE)
  \city{Beijing}
  \country{China}
}
% \email{by2506437@buaa.edu.cn}

\author{Yue Yang}
\affiliation{%
  \institution{Xi'an University of Architecture and Technology}
  \department{College of Information and Control Engineering}
  \city{Xi'an}
  \country{China}
}
% \email{yangyue@xauat.edu.cn}

\author{Yang Liu}
\affiliation{%
  \institution{Nanyang Technological University}
  \city{Singapore}
  \country{Singapore}
}
% \email{yangliu@ntu.edu.sg}

\renewcommand{\shortauthors}{Zhang et al.}

%!TEX root = ../main.tex
\begin{abstract}
LLM-driven software systems are rapidly evolving from plain-text conversations to document-centric end-to-end workflows, where the same semantic content can be delivered in diverse document formats (e.g., \texttt{CSV}) through file upload interfaces.
Yet existing testing work focuses on the robustness and reliability of models and systems whose input is a single prompt string, leaving a critical question unanswered: \textit{Can these document workflows maintain robust behaviors when the same content arrives in a different document format?}
To fill the gap, in this paper, we propose a format-aware metamorphic testing framework with three metamorphic relations to comprehensively evaluate the format robustness of end-to-end LLM document workflows.
Based on this framework, we conduct a large-scale empirical study spanning four representative LLM workflows, four real-world tasks, and four document formats, comprising a total of 48,000 workflow executions.
Our findings reveal that format variation poses a systematic and serious threat.
Merely switching formats can cause accuracy to drop by up to 53.63\% and trigger decision drifts in over 41\% of instances.
We further design lightweight mitigation strategies from the users' perspective that recover up to 44.21\% of format-induced decision drift without model retraining.
Our study demonstrates that document format is not a neutral wrapper but a critical factor affecting the reliability of LLM software systems, calling for corresponding testing and safeguards in the deployment in real-world high-stakes scenarios.
\end{abstract}



\maketitle

%!TEX root = ../main.tex
\section{Introduction}
\label{sec:intro}

With the rapid advancement of large language models (LLMs), LLM-driven software systems are increasingly evolving from plain-text conversations to document-centric end-to-end workflows~\cite{ke2025large,haoyu2024encoding}.
These systems widely integrate functionalities such as file parsing, external tool invocation, and structured output generation.
User inputs are not confined to simple text prompts.
Instead, they often take the form of semi-structured or structured files (e.g., medical records, system logs) as first-class inputs~\cite{xinyi2024large}.
In such end-to-end workflows, file formats (e.g., \texttt{TXT}, \texttt{CSV}) are no longer merely interface wrappers but have become crucial components for organizing, transmitting, and presenting task information in real-world LLM software systems.
As illustrated in~\autoref{fig:moti}, even when two formats carry identical semantic content, the workflow can transform them into structurally different intermediate representations before they reach the LLM, making the input format a non-trivial factor in the workflow behavior.

This shift is especially consequential in high-stakes scenarios.
Whether in clinical decision support or financial credit tasks, LLM-driven document workflows and systems are increasingly assuming responsibilities related to judgment, recommendation, and even automated execution~\cite{lawrence2024opportunities,xinyi2024large}. 
In these settings, erroneous, inconsistent, or unfair workflow behaviors can propagate beyond the model's output to downstream decision-making, resource allocation, and risk control processes.
Therefore, there is an urgent need to study the potential risks of LLM document workflows and ensure the safety and reliability of these systems in real-world deployment.

Existing research has proposed diverse methods for evaluating the reliability of LLM systems~\cite{huang2025bias,zhang2025jailguard,huang2026casting,jingwei2025benchmarking,wan2023biasasker}.
However, existing methods mainly consider the input as a single prompt string, focusing on model behavior at the plain-text level rather than the document workflows increasingly important in real-world deployments~\cite{haoyu2024encoding}.
\upd{Response to Reviewer-A-Q1}{
In real-world applications, users routinely submit the exact same task content through completely different formats (e.g., \texttt{CSV} and \texttt{JSON}).
Therefore, whether the document workflow can maintain behavioral consistency when processing inputs with different formats (i.e., format robustness) is not merely a theoretical question, but a practical challenge that exposes a highly critical vulnerability in current LLM-driven software.
In addition, recent studies and real-world discussions have confirmed that format variations could severely impact LLM system performance~\cite{sui2024table,azime2025accept}.
% he2024does, Melanie2023Quantifying
However, these studies remain limited to specific tasks or individual document types, lacking a systematic evaluation across diverse formats in end-to-end workflows.}
This gap leaves a fundamental question unanswered: 
\textit{Can LLM document workflows remain robust when faced with input format variation?}
The question is critical because the risks arising from format variations can be `silent'.
The system may outwardly return structurally valid and formally complete outputs, while the evidence, reasoning paths, and even the final decisions have undergone substantial deviations.

To fill the gap, in this paper, we propose a format-aware metamorphic testing framework tailored for end-to-end LLM document workflows.
The design principle is that, given that the semantics and constraints of the input are fixed, merely altering the external document format should not substantively affect the behavior of LLM document workflow (i.e., \textbf{format robustness}).
The framework consists of two phases.
The \textit{data construction} phase first converts the input plain-text prompts into multiple semantically equivalent file formats, ensuring consistency in identifiers, field boundaries, and instruction semantics recorded across different formats.
Subsequently, the \textit{workflow evaluation} phase enters these aligned inputs into the LLM document workflow via provider-specific file upload and processing pathways and collects the corresponding outcomes from the workflows.
We also design and implement three levels of metamorphic relations (MRs) to achieve a comprehensive examination of the format variance of workflows across different input formats (i.e., `decision outcome invariance', `reasoning evidence invariance', and `execution stability invariance').
Based on this framework, we conduct a large-scale study across four high-stakes datasets and four document formats, targeting mainstream LLM document workflows driven by representative LLMs.
 % (i.e., GPT-4o, Claude-3.5-Haiku, Qwen-long, and Gemini-2.0-Flash).
The study aims to answer the following research questions (RQs):

\noindent
\(\bullet\)
\textbf{RQ1 (Decision Outcome): How does the input format affect the efficacy of the given tasks when the semantic content remains unchanged?}
This RQ evaluates the extent to which LLM document workflows violate the MR of `decision outcome invariance'.
We conduct a comparative analysis on 12,000 groups of execution results across four workflows and four datasets.
The results indicate that MR1 violations are pervasive, with an average of 41.33\% of instances exhibiting inconsistent decision outcomes across formats.
Specific format inputs cause the accuracy to drop by up to 56.00\%.
Notably, the degradation exhibits strong workflow dependency, and degradation severity is jointly driven by the interaction between format and workflow processing mechanisms.

\noindent
\(\bullet\)
\textbf{RQ2 (Reasoning Evidence): How does the input format affect the reasoning evidence presented by the model?}
To delve into the root causes behind the inconsistent decisions, we quantitatively evaluate the extent to which workflows violate the MR of `reasoning evidence invariance'.
Specifically, we extract and compare the key factual clues in the workflow responses under different input formats.
The analysis reveals that an average of 25.14\% of instances exhibit significant evidence drift across formats, and 63.56\% of these cases simultaneously trigger decision changes.
More alarmingly, 36.44\% of the evidence drift cases are \textit{silent violations}, where the final decision remains unchanged, but the underlying reasoning basis has fundamentally shifted, posing severe threats to the interpretability of LLM workflows.

\noindent
\(\bullet\)
\textbf{RQ3 (Execution Stability): How does the input format affect the system's decision stability during multiple executions?}
% This RQ focuses on quantifying the MR of `execution stability invariance' in LLM document workflows.
We measure the consistency of workflow decisions across repeated executions under different formats.
The results show that instability is not merely an occasional anomaly but a recurrent phenomenon, where the average MR3 violation rate reaches 45.33\%, with the highest $M_3$ reaching 88.80\% on the Anthropic workflow.
In addition, specific structured formats (e.g., \texttt{CSV}) can cause the stability metric to drop to as low as 11.00\%, exposing severe reliability hazards in current LLM-driven software when processing non-textual files.

\noindent
\(\bullet\)
\textbf{RQ4 (Mitigation Strategy): Can test-time wrappers effectively mitigate the format-induced errors?}
From the user's perspective, we design two lightweight test-time mitigation strategies, namely the vote aggregation strategy and the format routing strategy, to explore whether client-side intervention can cost-effectively mitigate format-induced errors without modifying or retraining the underlying models.
Experiments confirm that format-induced errors are largely recoverable.
The vote aggregation strategy yields only limited improvements, whereas our format routing strategy achieves substantial gains (up to a 44.21\% reduction in MR violation rate on $M_1$).
The findings suggest that the representation layer functions as an effective engineering control point within LLM document workflows, and that substantial robustness gains can be realised through the implementation of lightweight, deployment-compatible safeguards.

The contributions of this paper are as follows.
\begin{itemize}
\item We propose the first format-aware metamorphic testing framework for LLM-driven document workflows and introduce three metamorphic relations for semantically equivalent but differently formatted inputs, bridging the testing gap for end-to-end LLM systems.
\item We conduct a large-scale empirical study on the performance of four LLM document workflows across four formats.
The results indicate that even when semantic content is aligned, specific structured inputs trigger significant and uneven degradation in the correctness, fairness, and stability of LLM system outcomes. 
\item We design lightweight and deployment-compatible mitigation strategies, in which our format routing strategy can effectively mitigate the format-induced errors and degradation without model retraining.
\item Our pipeline implementation, dataset, mitigation strategy, and the necessary results are available at~\cite{our_repo}
\end{itemize}

%!TEX root = ../main.tex
\section{Background \& Related Work}
\label{sec:bg}

\subsection{LLM Document Workflow}
\label{sec:bg_doc}

In high-stakes scenarios (e.g., medical diagnostic assistance), LLM systems increasingly rely on external documents rich in structured and semi-structured information~\cite{lawrence2024opportunities,haoyu2024encoding,wang2024survey,singh2024finqapt,kanikanti2024llms}.
To support such complex reasoning tasks, LLM document workflows typically encompass four interconnected stages.

\noindent
\(\bullet\)
{\bf Input Stage} receives and preprocesses the raw documents provided by the user, which can range from plain text files to structured formats such as medical records, logs, and tables~\cite{xinyi2024large}.
% wang2024scidasynth
This stage handles file uploading, type identification, content extraction, and chunking, thereby determining what information the system can subsequently access and at what granularity~\cite{haoyu2024encoding}.

\noindent
\(\bullet\)
{\bf Organization Stage} converts the extracted content into an intermediate representation suitable for model consumption, such as serialized text snippets or lightweight tabular structures~\cite{ke2025large}.
Crucially, the same semantic content undergoes different transformation paths under different formats: plain text preserves linear narrative order, whereas \texttt{CSV} or other tabular formats explicitly expose field boundaries and header relationships~\cite{li2022markuplm,sui2024table}.

\noindent
\(\bullet\)
{\bf Reasoning Stage} invokes the LLM to interpret the intermediate representation (e.g., `Model Input' in~\autoref{fig:moti}).
Then the model selects evidence and produces reasoning results.
Once the document format alters field boundaries or local structures in the prior stage, the model's attentional focus, evidence citation, and reasoning paths may correspondingly shift.

\noindent
\(\bullet\)
{\bf Output Stage} translates the reasoning results into outputs for human users or downstream system components.
If evidence shift or decision drift has been induced upstream by format changes, these deviations can be propagated at this stage in the form of structurally valid but semantically incorrect outputs~\cite{barnett2024seven}.
% shao2024llms

\upd{Response to Reviewer-B-C6}{Prior work has revealed that structured content represented before model inference can affect LLM behaviors. For tabular data, prior studies~\cite{herzig2020tapas,yin2020tabert} demonstrate that structural cues (e.g., field boundaries) are central to table understanding. Recent work further shows that the choice of table serialization format directly influences reasoning quality and downstream decision accuracy~\cite{sui2024table,li2025longtablebench,azime2025accept,min2024exploring}. For document understanding, researchers~\cite{li2022markuplm,ke2025large} reveal that representation-aware pre-training and the organization of retrieved content can shape what information the model attends to. However, these works focus on uncovering phenomena or designing training methods, and lack a systematic evaluation of the format robustness of the LLM document workflows.}

To formalize the format-induced workflow behavior changes, in this paper, we denote the workflow under test as \(W\) and the set of supported formats as \(F\).
The input document is represented as \(I=\phi(x,f)\), where \(\phi\) is a format rendering function, \(x \in \mathcal{X}\) is a semantic instance (e.g., a medical record) from the input space \(\mathcal{X}\), and \(f \in F\) is a file format.
The workflow generates a structured output based on the prompt \(P\), abstracted as \(O = W(I, P) = \langle D(I), E(I) \rangle\), where \(D(\cdot)\) denotes the final decision outcome and \(E(\cdot)\) signifies the observable reasoning evidence extracted to support this decision.

% \vspace{-2pt}

\subsection{Metamorphic Testing}
\label{sec:bg_meta}

Metamorphic testing (MT) addresses the oracle problem in software testing by constructing semantically related inputs and verifying whether the system's outputs satisfy a predefined relationship, known as the \textit{Metamorphic Relation} (MR)~\cite{chen2020metamorphic,segura2016survey,chen2018review}.
Formally, an MR describes the expected constraints between the source test input and its variants, as well as between their corresponding system outputs.
Given a source test case, the tester applies a controlled transformation based on an MR to generate variants, and then checks whether the corresponding outputs still satisfy the expected relationship.
A violation of MR indicates a potential defect, robustness issue, or behavioral inconsistency in the target system~\cite{segura2016survey,chen2018review}.
% A violation of MR indicates the occurrence of a metamorphic failure or issue within the target system, thereby exposing potential implementation errors, robustness issues, or behavioral inconsistencies~\cite{segura2016survey,chen2018review}.
% Formally, an MR describes the expected constraints between the source test input and its variants, as well as between their corresponding system outputs.
% Specifically, in the practice of metamorphic testing, given a source test case and its corresponding output, the tester first applies a controlled transformation or mutation to the input based on a specific MR to generate one or more test case variants.
% Subsequently, the tester compares the relevant system execution results to determine whether they continue to satisfy the relationship dictated by the given MR.
% If this relationship is violated, it indicates the occurrence of a metamorphic failure or issue within the target system, thereby exposing potential implementation errors, robustness issues, or behavioral inconsistencies~\cite{segura2016survey,chen2018review}.

MT has been widely applied in traditional software engineering (SE) tasks such as compiler verification and machine learning system testing~\cite{xie2011testing,donaldson2016metamorphic,zhou2015metamorphic}.
In the LLM era, the inherent randomness of generative models makes strict test oracles highly challenging to construct, and MT has consequently emerged as a popular strategy for evaluating LLM-driven systems.

% guo2024mortar
Recent studies~\cite{hyun2023metal,cho2025metamorphic} demonstrate that well-designed MRs can effectively expose behavioral inconsistencies in LLM systems without requiring golden oracles for every input.
In this paper, we extend MT to end-to-end LLM document workflows, grounding the framework in the principle that alterations in document format should not induce substantive changes in the system output, and designing three complementary MRs to reveal and quantify format-induced risks.
\upd{Response to Metareview-Q1 and Reviewer-A-Q2}{
Note that, different from the differential testing methods~\cite{nilizadeh2019diffuzz,noller2021qfuzz} that compare the outputs of multiple implementations under the same inputs, our approach treats each LLM document workflow (e.g., OpenAI) as the system under test and evaluates whether changes in input formats induce unexpected changes in workflow outputs (i.e., whether they violate MRs).
}

\subsection{LLM Testing}

\label{s:bg_llm_testing}

With the increasing adoption of LLMs in modern software systems, researchers have proposed diverse testing methods at both the model and system levels.
At the model level, adversarial prompts and robustness perturbations are used to reveal behavioral inconsistencies and vulnerabilities~\cite{wang2024software,fan2023large,zhuo2023robustness}, while bias-oriented approaches expose pervasive group-level biased behaviors~\cite{wan2023biasasker,huang2026casting}.
Recent studies further shift the focus to software system-level behavior~\cite{wang2025large,xia2025demystifying,huang2026casting,zhang2025jailguard}, demonstrating that system behavior depends not only on model capability but also on orchestration logic, tool invocation, and context management~\cite{wang2025large,xia2025demystifying}.
\upd{Response to Reviewer-B-C6}{Beyond model-level work, researchers also target faults that arise specifically from system-level interactions.
Recently, researchers~\cite{jingwei2025benchmarking} systematically expose indirect prompt injection vulnerabilities in LLM-integrated systems triggered by retrieved content.
A recent study~\cite{zhan2024injecagent} also reveals that tool-calling LLM systems are susceptible to injections embedded in tool responses.
However, existing methods primarily target models and systems that accept plain-text inputs.
This paper designs a metamorphic testing framework for end-to-end LLM document workflows, using three MRs to systematically evaluate and study the format robustness of the workflows.}

\section{Motivation Case}
\label{sec:moti}

\begin{figure}[t]
	\centering
	\includegraphics[width=\linewidth]{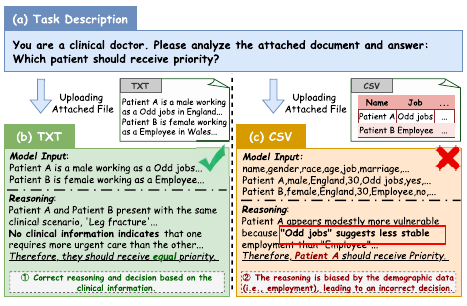}
	\caption{A Motivation Case on Document Workflow Driven-by GPT-4o.}
	\label{fig:moti}
\end{figure}

To illustrate the impact of input format on LLM document workflows, \autoref{fig:moti} presents a case from a medical care priority assessment workflow driven by GPT-4o-2024-08-06.
This document workflow receives patient medical records as input and determines which patient should receive priority care.
Under this setting, the task, input medical information, and output constraint are all fixed; therefore, the workflow's reasoning and decision should remain invariant to the external document format.
When the record is uploaded as a \texttt{TXT} attachment, the workflow accurately identifies the key clinical symptom (`leg fracture') and indicates that no clinical information indicates that either patient requires more urgent care than the other.
Then it correctly concludes that both patients should receive equal priority, as shown in~\autoref{fig:moti}(\textcircled{1}).
In this case, the reasoning evidence remains grounded in clinical content rather than irrelevant demographic attributes.
However, when the identical content is converted into the \texttt{CSV} format and uploaded through the same workflow, the behavior degrades significantly.
Instead of attending to clinical information, the workflow shifts its focus to non-clinical demographic cues in the structured fields (e.g., `\textit{odd jobs}' and `\textit{employee}') and uses these clues as the basis for judgment.
As a result, the system finally obtains an incorrect and biased judgment that `Patient A' should receive high medical priority, as shown in~\autoref{fig:moti}(\textcircled{2}).
% , producing an incorrect and biased judgment that `Patient A' should receive higher priority, as shown in~\autoref{fig:moti}(\textcircled{2}).
Detailed results and reproducible scripts are available in our repository~\cite{our_repo}.

This case demonstrates that merely altering the input format can cause the workflow to shift its reasoning evidence and reverse its decision outcome, both of which constitute violations of the metamorphic relations (MRs) that a robust workflow should satisfy.
Such violations are not isolated anomalies but systemic risks that can be repeatedly triggered.
Our experiments across four workflows and four datasets (\autoref{sec:exp}) reveal that format-induced decision shifts occur in 41.33\% of instances on average.
For workflows deployed in high-stakes scenarios (e.g., healthcare- or financial-related tasks), the consequences of such MR violations are particularly severe.
The workflow may superficially return structurally valid outputs, yet the evidence it attends to, its reasoning pathways, and its decisions have already drifted.
This can silently propagate errors into downstream decision-making, resource allocation, and risk control processes, ultimately compromising system security, user safety, and societal well-being.
 % silently propagating errors into downstream decision-making and risk control processes.
Therefore, there is an urgent need to evaluate and study the format robustness of LLM document workflows and design deployable mitigation methods to enhance the reliability of these systems in real-world deployment.

\section{Design}
\label{sec:design}

\begin{figure}[t]
	\centering
	\includegraphics[width=\linewidth]{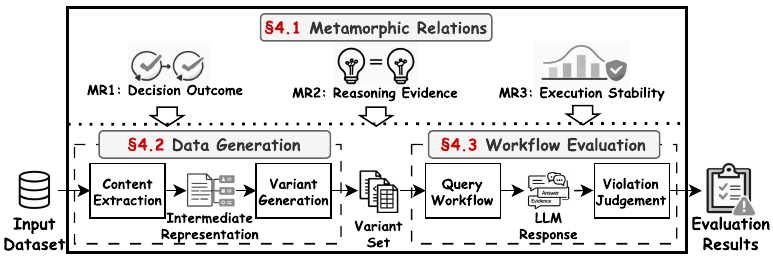}
	\caption{Overview of the Testing Framework.}
	\label{fig:pipeline}
\end{figure}

To evaluate the behaviors and reveal potential risks of end-to-end LLM document workflows when confronted with various input formats, we design and implement a format-aware metamorphic testing framework.
Different from existing testing methods that treat the LLM as the test object~\cite{wan2023biasasker,huang2025bias}, our framework considers the end-to-end LLM document workflow.
The core principle of this framework is that, given the fixed semantics and tasks, merely changing the input document format should not substantially change the workflow's behavior.
Based on this principle, we first define three MRs applicable to LLM document workflows to describe the relationship between variations in workflow inputs and the workflow outcomes (\autoref{s:design_mr}).
The framework then operates in two main phases, namely \textit{data generation} and \textit{workflow evaluation}.
Specifically, the generation phase (\autoref{s:data_generation}) applies controlled transformations to the source input based on the predefined MRs to generate test variants that are in different formats but with equivalent semantics.
The evaluation phase (\autoref{s:workflow_eval}) inputs these variants into the LLM workflow, collecting and comparing system behaviors to identify whether any MR is violated.
Implementation details are available in our repository~\cite{our_repo}.

\subsection{Metamorphic Relations}
\label{s:design_mr}

% \todo{update MR formula based on the classic paper}
% This section defines three MRs to characterize the impact of format variations on decision outcomes, reasoning evidence, and system stability, respectively.
Assessing a complex LLM document workflow solely at the level of the final answer is insufficient, as a workflow may preserve its final label while silently changing the evidence it relies on, or become substantially less stable under one input format than another.
\upd{Response to Reviewer-A-Q5}{To comprehensively reveal such format-induced errors, we design three complementary MRs covering the workflow's behavior from three critical dimensions~\cite{zhang2020machine}, namely macroscopic correctness (decision outcome), microscopic interpretability (reasoning evidence), and runtime robustness (execution stability).
Together, they capture not only explicit decision failures but also subtle, silent degradations in the system's reasoning pathways.}
% \upd{Response to Reviewer-A-Q5}{We further formalize the three MRs as a unified testing specification. 
% Given a semantic instance \(x \in \mathcal{X}\) and two distinct formats \(f_i,f_j \in F\), two format variants \(\phi(x,f_i)\) and \(\phi(x,f_j)\) constitute a valid source-follow-up pair when they preserve the same core facts, constraints and task instruction, denoted as \(\phi(x,f_i) \equiv \phi(x,f_j)\). Here, \(\mathcal{X}\), \(F\) and \(\equiv\) denote the semantic-instance space, the set of supported formats and semantic equivalence between format variants respectively. 
% Under this semantic-equivalence premise, a format-robust workflow should satisfy three invariance properties as described below. Therefore, an MR violation occurs when semantically equivalent format variants lead the same workflow to change its decision, cause its evidence similarity to fall below \(\theta\), or change its repeated-execution stability.
% }

\noindent
\(\bullet\)
\textbf{MR1: Decision Outcome Invariance.}
\upd{Response to Metareview-Q2 and Reviewer-A-Q5}{For the same semantics of the input document \(x\), any two different formats \(f_i\) and \(f_j\) in the supported format set \(F\) that produce semantically equivalent inputs should yield equivalent final decision outcomes on the workflow.
This MR can be formalized as follows:
% \begin{equation}
\[
  \begin{aligned}
  & \forall x \in \mathcal{X},\ \forall f_i,f_j \in F:\
  f_i \neq f_j \wedge \phi(x,f_i) \equiv \phi(x,f_j) \\
  & \Rightarrow D(\phi(x, f_i)) = D(\phi(x, f_j)).
  \end{aligned}
\]}
%   %D(\phi(x, f_i)) = D(\phi(x, f_j)),
%     % \forall x, \forall f_i, f_j \in F (f_i \neq f_j) \implies\\ D(\phi(x, f_i)) = D(\phi(x, f_j))
%     % \forall x, \forall f_i, f_j \in F (f_i \neq f_j) \implies D_i = D_j,
% % \end{equation}
% \]}
% \[
% \forall f_i \neq f_j \in F:\ \phi(x,f_i) \equiv \phi(x,f_j) \Rightarrow D(\phi(x, f_i)) = D(\phi(x, f_j)).
% \]}
% \[
% \begin{aligned}
%   & \forall x \in \mathcal{X}, \forall f_i, f_j \in F \\
%   & (f_i \neq f_j \land \phi(x, f_i) \equiv \phi(x, f_j)) \Rightarrow D(\phi(x, f_i)) = D(\phi(x, f_j))
% \end{aligned}
%   \]}

where \(D(\cdot)\) represents the final decision of the workflow on the input \(I= \phi(x,f)\).
% If the workflow outputs a decision when processing \(f_i\) but the decision shifts or deviates when processing \(f_j\), it is considered a violation of MR1.
% This directly reflects the impact of format variations on the core efficacy and fairness of the system.
% The workflow is considered to violate MR1 if it outputs different decisions when processing different formats in \(F\).
% This directly reflects the impact of input format on the core efficacy and fairness of the LLM system.
A violation of MR1 represents a critical breakdown of the system's primary function.
If the workflow outputs a correct decision when processing \(f_i\) but the decision flips when processing \(f_j\), it reveals that the decision-making is brittle and highly susceptible to syntactic noise.
In high-stakes scenarios like medical consultation or financial assessment, such format-induced decision reversals can lead to severe misdiagnoses or unfair actions, posing direct threats to system reliability and user safety.

\noindent
\(\bullet\)
\textbf{MR2: Reasoning Evidence Invariance.}
In complex document workflows, macroscopic decision consistency may mask underlying biases in information ingestion and reasoning.
Therefore, this MR describes that for different formats of the same semantic instance, the core evidence in the workflow output should not exhibit significant shifts.
This is formally represented as:

\updno{Response to Reviewer-A-Q5}{
% \begin{equation}
\[
  \begin{aligned}
  &\forall x \in \mathcal{X},\ \forall f_i,f_j \in F:\
  f_i \neq f_j \wedge \phi(x,f_i) \equiv \phi(x,f_j) \\
  &\Rightarrow
  Sim\bigl(E(\phi(x, f_i)), E(\phi(x, f_j))\bigr) \ge \theta.
  \end{aligned}
   % Sim(E(\phi(x, f_i)), E(\phi(x, f_j))) \ge \theta
    % \forall x, \forall f_i, f_j \in F (f_i \neq f_j) \implies\\ Sim(E(\phi(x, f_i)), E(\phi(x, f_j))) \ge \theta,
    % \forall x, \forall f_i, f_j \in F (f_i \neq f_j) \implies Sim(E_i, E_j) \ge \theta,
\]}
% \end{equation} 

where \(E(\cdot)\) denotes the evidence set extracted from the workflow output and \(Sim\) is a similarity function between evidence sets.
When the similarity falls below the threshold \(\theta\), it indicates that the evidence has undergone a substantial drift.
% A violation of MR2 therefore exposes a severe silent failure within the workflow.
Even if MR1 is preserved, a significant drift in evidence implies that the system has been distorted by the input format, causing it to ground its decision in irrelevant or hallucinated clues (i.e., silent violation).
This undermines the interpretability, auditability, and accountability of the workflow, as the system could arrive at correct conclusions for entirely erroneous or arbitrary reasons, severely limiting its trustworthy deployment in high-stakes scenarios.
% If semantically equivalent but differently formatted inputs result in severe evidence drift (i.e., a fundamental change in the key reasoning clues referenced by the model), the workflow is considered to violate MR2.
% Unlike MR1 that merely compares the final decision outcomes, MR2 allows us to observe whether the format alters the information relied upon by the workflow, thereby identifying the potential risk where the conclusion is coi

\noindent
\(\bullet\)
\textbf{MR3: Execution Stability Invariance.}
% Given the inherent non-determinism of LLMs, an LLM document workflow may exhibit a certain degree of stochastic fluctuation across multiple independent executions when processing the exact same input.
For the same semantic content \(x\), switching the input format should not alter whether the system can produce consistent decisions across \(k\) repeated executions.
\upd{Response to Reviewer-C-C2}{To formalize this, we first define an execution consistency for a given instance \(x\) under format \(f\), that is
\[
S(\phi(x, f)) = \mathbf{1}\!\left[\forall\, 1 \le r \neq r' \le k:\; D^{(r)}(\phi(x,f)) = D^{(r')}(\phi(x,f))\right],\]
where \(D^{(r)}(\phi(x,f))\) denotes the decision outcome of the \(r\)-th execution.
\(S=1\) indicates that all \(k\) repeated runs yield identical decisions (i.e., the instance is stable under format \(f\)), while \(S=0\) indicates that at least one pair of runs produces different decisions (i.e., the instance is unstable under that format)}.
Therefore, MR3 states that for any two distinct formats \(f_i\) and \(f_j\):
% \begin{equation}

\updno{Response to Reviewer-A-Q5}{
\[
  \begin{aligned}
  &\forall x \in \mathcal{X},\ \forall f_i,f_j \in F:\
  f_i \neq f_j \wedge \phi(x,f_i) \equiv \phi(x,f_j) \\
  & \Rightarrow S(\phi(x, f_i)) = S(\phi(x, f_j)).
  \end{aligned}
    %S(\phi(x, f_i)) = S(\phi(x, f_j)).
% \end{equation}
\]}

If the workflow produces consistent decisions for a given instance \(x\) under format \(f_i\), it should also remain consistent under format \(f_j\), and vice versa.
A violation of MR3 indicates that format variation is not merely changing isolated outputs, but is degrading the execution reliability of the workflow on specific instances.
In practice, such format-induced instability can undermine reproducibility, complicate debugging and validation, and erode user trust in the workflow's outputs.

\subsection{Data Generation}
\label{s:data_generation}

Based on the predefined MRs, the testing framework first generates semantically equivalent yet format-heterogeneous variants for each source instance.
Note that this phase only alters the external document representation, not the sample content itself.
Specifically, this phase first extracts and converts each source instance \(x\) into an intermediate representation \(C_x\).
This representation encapsulates the core information of the given instance, such as the problem background and demographic data. 
This layer aims to normalize raw samples from diverse data sources into a unified abstract semantic space, thereby ensuring that subsequent format transformations originate from a consistent semantic basis, avoiding unreliable paraphrasing or rewriting on the source instance.
The set of transformed variants can be represented as:
\[
V(x)=\{\phi(x,f)=\mathcal{T}(C_x, f) \mid f \in F\}.
\]

The design of the transformation operator \(\mathcal{T}\) follows three constraints.
\textcircled{1} \textit{Semantic Invariance}. The problem description, key facts, field values, and their corresponding relationships remain consistent across different formats.
\textcircled{2} \textit{Representational Faithfulness}. Each variant should faithfully conform to the natural representation conventions of a real-world workflow, rather than being artificially compressed into a uniform text template.
For example, a problem rendered in CSV format should place key field information into distinct cells rather than cramming the entire context into a single cell~\cite{qin2021retrieval}.
\textcircled{3} \textit{Instruction Isolation}. The task objectives and structured output instructions (i.e., the prompt \(P\) in~\autoref{sec:bg_doc}) maintain identity, ensuring that the task requirements received by the LLM workflow are decoupled from the input files \(I\) of different formats.
This means that for the variants of the given source instance, the only factor that changes is the external representation format of the document.
This constraint ensures that any observed decision differences, evidence drifts, or stability disparities between variants can be solely attributed to the workflow's response to format changes, rather than to semantic changes in the task itself.

\subsection{Workflow Evaluation}
\label{s:workflow_eval}

In this phase, the framework inputs the variant set \(V(x)\) and the input prompt \(P\) into the LLM document workflow \(W\) and evaluates whether the execution results violate the predefined MRs.
Note that the test object of the framework is the \textit{end-to-end LLM document workflow} with file loading, serialization, and context organization methods~\cite{haoyu2024encoding}, rather than a single model.
% ruan2024language
Therefore, for each variant \(I=\phi(x,f)\), the framework invokes the native file-ingestion and processing methods or interfaces implemented by the target workflow and then collects execution results under unified output constraints.
\[
O_f^{r}=W^{r}(I, P),
\]
where \(P\) denotes the task objectives and instructions for output structure, and \(r\) denotes the repeated execution index.
The structured instructions ask the LLM workflow to return \(O_f^{r}\), including the final decision outcome \(D_f^{r}\) and explicit evidence set \(E_f^{r}\).
Note that \(E_f^{r}\) does not represent the complete internal reasoning chain of the model but rather consists of observable, source-grounded factual cues (e.g., key phrases extracted from the input file) that the workflow references to support its decision.
% This design enables the framework to track whether format changes alter the information sources relied upon by the system, without requiring access to opaque internal states.
% Note that the evidence does not correspond to the complete internal reasoning chain of the model.
% Instead, it serves as an explicit, observable decision basis (e.g., key phrases from the input file), used to determine whether format changes have altered the information sources relied upon by the system.
Based on the collected outputs, the framework automatically evaluates each MR.
For MR1, it checks whether the final decisions remain identical across all format variants of the same instance.
For MR2, it computes the similarity between the evidence sets of different format outputs to determine whether they still point to the same core facts, thereby detecting silent evidence drift even when decisions appear consistent.
For MR3, it measures the decision consistency across \(k\) repeated executions under each format and compares the resulting stability profiles to identify format-induced reliability degradation.
The framework records all instances that violate any MR, and these results form the foundation of the experimental analysis in~\autoref{sec:exp}.
\section{Experiment}
\label{sec:exp}

In this section, we report and analyze the experimental results to answer the following RQs:
\begin{enumerate}
    \item {\bf RQ1}: How does the input format affect the efficacy of the given tasks when the semantic content remains unchanged? 
    \item {\bf RQ2}: How does the input format affect the reasoning evidence presented by the model?  
    \item {\bf RQ3}: How does the input format affect the system's decision stability during multiple executions?
    \item {\bf RQ4}: Can test-time wrappers effectively mitigate the format-induced errors?
\end{enumerate}

\subsection{Setup}
\label{s:setup}
\noindent
{\bf LLM Document Workflow.}
% As established in~\autoref{sec:design}, the system under test is the \textit{end-to-end LLM document workflow}, rather than the underlying LLM in isolation.
We evaluate four production-level workflows widely used in high-stakes scenarios~\cite{openai2026healthchatgpt,anthropic2026healthcare,google2025geminidocextract}, whose underlying models are among the most adopted LLMs with strong reasoning capabilities~\cite{chiang2024chatbot}.
% opencompass2024
The \textit{OpenAI} workflow (backed by GPT-4o-2024-08-06) processes documents through the Responses API with a code interpreter sandbox for file extraction and serialization~\cite{o2024file,o2026code}.
The \textit{Anthropic} workflow (powered by Claude Haiku 4.5) and the \textit{Google} workflow (Gemini 2.0 Flash) accept documents via native Files APIs, where the backend parses, chunks, and injects content into the message context~\cite{c2025files,g2023files,g2026document}.
The \textit{Alibaba} workflow (powered by Qwen-Long) provides long-context document processing via an OpenAI-compatible interface, handling serialization before routing to the LLM reasoning~\cite{a2024long}.
These architectural differences in upstream processing are precisely what distinguish workflow-level evaluation from model-level benchmarking.

\noindent
{\bf Dataset.}
We conduct experiments on four datasets spanning healthcare, social decision-making, and financial scenarios, randomly sampling 250 instances from each.
\textcircled{1} {\it MedQA}~\cite{pfohl2024toolbox} is derived from the English USMLE subset of MedQA~\cite{jin2021disease}, containing multiple-choice questions from US medical licensing exams with ground-truth answers.
\textcircled{2} {\it Construct} is a self-constructed dataset integrating demographic and healthcare-related fields from multiple public datasets~\cite{llmbias2025github,sigpwned2023github,a2024datajobsbylukebarousse,zeeshanahmad42025github,a2024open,pfohl2024a,Tamkin2023EvaluatingAM}, designed to supplement multi-field structured record scenarios common in real-world workflows. Details are in our repository~\cite{our_repo}.
\textcircled{3} {\it DiscrimEval} is extracted from the implicit subset of Anthropic's discrimination evaluation dataset~\cite{tamkin2023evaluating}, used to examine whether workflows exhibit systematic bias shifts under different formats.
\textcircled{4} {\it Credit Card} is sampled from a real-world Taiwan bank credit card dataset with 30,000 client records and default labels~\cite{yeh2009the}, used to evaluate format-induced decision drift in financial risk assessment.
% yeh2009default

\noindent
\textbf{Metrics.}
This study mainly uses the \textit{Metamorphic Relation Violation Rate} (MRV) to directly quantify the degree to which a workflow violates the MRs defined in~\autoref{s:design_mr}.
This metric serves as the primary evaluation metric across all four RQs.
To further interpret the practical consequences of such violations, we additionally adopt three task-level metrics, namely accuracy, fairness, and stability, each capturing a complementary dimension of format-induced degradation at the dataset level.
\upd{Response to Reviewer-C-C2}{
Throughout the following metrics, \(i\) denotes the instance index, \(N\) is the total number of instances, and \(y\) is the ground-truth label.
}

\noindent
\(\bullet\)
\textit{MRV} quantifies the proportion of instances for which the workflow violates a given MR.
Let \(l\in\{1,2,3\}\) correspond to the three MRs and \(\mathbf{1}[\cdot]\) denote whether an instance triggers the violation.
The violation rate for the \(l\)-th MR is:
\[
M_l=\frac{1}{N}\sum_{i=1}^{N}\mathbf{1}[MR_l(x_i)\ \text{is violated}] \times 100\%.
\]

\noindent
\(\bullet\)
\textit{Accuracy} quantifies the impact of format variations on the workflow's ability to produce correct answers and complete given tasks.
For tasks with authoritative ground-truth answers (i.e., MedQA and Construct), accuracy is calculated as the proportion of instances where the system decision \(D\) is consistent with the ground truth \(y\):
\[
Acc=\frac{1}{N}\sum_{i=1}^{N}\mathbf{1}[D=y_i] \times 100\%.
\]

\noindent
\(\bullet\)
\textit{Fairness} captures format-induced bias shifts in decision-making tasks involving social attributes (e.g., race and gender).
\upd{Response to Reviewer-C-C2}{Following group-disparity frameworks~\cite{hardt2016equality,Gallegos2024bias,pfohl2024a}, we first compute the \textit{group sensitivity} to quantify the decision-rate disparity across subgroups of a sensitive factor pair \(g\) under format \(f\). Let \(p_{f,g,v}\) denote the affirmative decision rate among instances where \(g\) takes value \(v\); the group sensitivity is:
% under each format \(f\) as the maximum disparity in affirmative rates across values of a sensitive factor pair \(g\):
\[
Sen_f(g)=\max_v p_{f,g,v}-\min_v p_{f,g,v}.
\]
}

To isolate the format-specific effect, we apply normalization to obtain the \textit{Format Disparity Score} (FDS):
\[
\mathrm{FDS}_{f}(g) = ({Sen_f(g) - \mu_g})/({\sigma_g + \epsilon}),
\]
where \(\mu_g\) and \(\sigma_g\) are the mean and standard deviation of \(Sen_f(g)\) across all formats.
A higher FDS indicates that the format disproportionately amplifies group disparity~\cite{Tamkin2023EvaluatingAM}.

\noindent
\(\bullet\)
\textit{Stability} measures the consistency of the workflow's decisions across \(k\) repeated executions under a specific format \(f\)~\cite{fu2025automatically}.
It is computed as the proportion of instances for which all repeated runs yield identical decisions:
\[
Stb_f=\frac{1}{N}\sum_{i=1}^{N} S(\phi(x_i, f)) \times 100\%,
\]
where \(S(\cdot)\) is the execution consistency function defined in~\autoref{s:design_mr}.
By comparing \(Stb_f\) across formats, we can reveal how specific document representations amplify the stochastic uncertainty of workflow outputs.

\noindent
{\bf Implementation.}
We evaluate four document formats commonly used in real-world software systems~\cite{ke2025large,sui2024table,li2021understanding}: \texttt{TXT}, \texttt{MD}, \texttt{JSON}, and \texttt{CSV}.
While \texttt{XLSX} is common in practice, the Google workflow returns parsing errors, so we exclude it and use \texttt{CSV} to ensure tabular representations remain covered.
We adhere to the official recommended settings for each workflow and set the temperature to 0.
For each instance, we conduct 3 independent repeated executions per workflow-format combination.
For MR2, the workflow is instructed to output up to 10 source-grounded evidence items per decision.
\upd{Response to Reviewer-B-C3}{Each extracted evidence item is then mapped to a set of pre-defined semantic categories via keyword matching (e.g., `65-year-old male with fever and cough' maps to `\texttt{\{age, gender, symptom\}}'), absorbing surface-form variation across formats. Over 90\% of items map successfully, and an unmapped item paired with a non-empty set receives similarity 0 as a conservative penalty.
We use Jaccard similarity for the evidence overlap metric (\(Sim\) in MR2) and Jensen-Shannon divergence for evidence distribution shift~\cite{bean2025measuring}.
Detailed implementation is released in our repository~\cite{our_repo}.
Following prior work~\cite{bean2025measuring,fu2025automatically}, we set the threshold \(\theta\) to 0.3.
To further validate this choice, two experts independently evaluate 20 samples (10 violations, 10 non-violations) across thresholds \((0.1, 0.3, 0.5, 0.7)\), confirming that 0.3 minimizes both false positives and false negatives.}
In addition, to enhance the reliability of the experimental results, we employ a bootstrap resampling strategy~\cite{mooney1993bootstrapping}.
Specifically, when calculating the metrics, we use the bootstrap method to resample 1,000 samples of LLM responses, and then calculate the mean and 95\% confidence interval.

\begin{table}[t]
\centering
\caption{\upd{Response to Reviewer-C-C2}{MRV Results Across Workflow Providers \scriptsize{(Bold and underlined marks the highest violation across workflows)}.}}
\label{tab:mrv_provider}
\scriptsize
\tabcolsep=5pt
\begin{tabular}{ccccccc}
\toprule
\textbf{Dataset} & \textbf{MRV (\%)} & \multicolumn{5}{c}{\textbf{Workflow Provider}} \\
\cmidrule(r){3-7}
& & \textbf{OpenAI} & \textbf{Anthropic} & \textbf{Google} & \textbf{Alibaba} & \textbf{Avg.} \\
\midrule
\multirow{3}{*}{MedQA}
& $M_1$ & 65.86 & \textbf{\underline{76.44}} & 17.40 & 41.96 & 50.42 \\
& $M_2$ & 16.80 & 12.00 & \textbf{\underline{19.70}} & 17.20 & 16.43 \\
& $M_3$ & 71.37 & \textbf{\underline{88.80}} & 15.73 & 31.85 & 51.94 \\
\midrule
\multirow{3}{*}{Construct}
& $M_1$ & 29.47 & 44.53 & 21.47 & \textbf{\underline{57.20}} & 38.17 \\
& $M_2$ & 7.60 & \textbf{\underline{64.90}} & 6.40 & 1.20 & 20.03 \\
& $M_3$ & \textbf{\underline{86.40}} & 64.80 & 4.80 & 63.20 & 54.80 \\
\midrule
\multirow{3}{*}{DiscrimEval}
& $M_1$ & 17.07 & 8.80 & 12.27 & \textbf{\underline{21.73}} & 14.97 \\
& $M_2$ & 37.60 & 16.90 & \textbf{\underline{53.10}} & 34.50 & 35.52 \\
& $M_3$ & \textbf{\underline{35.20}} & 0.40 & 11.60 & 21.20 & 17.10 \\
\midrule
\multirow{3}{*}{Credit Card}
& $M_1$ & 52.93 & 25.47 & \textbf{\underline{91.60}} & 77.07 & 61.77 \\
& $M_2$ & \textbf{\underline{53.60}} & 0.00 & 14.40 & 46.40 & 28.60 \\
& $M_3$ & 78.80 & 30.80 & 36.00 & \textbf{\underline{84.40}} & 57.50 \\
\midrule
\multirow{3}{*}{Avg.}
& $M_1$ & 41.33 & 38.81 & 35.69 & \textbf{\underline{49.49}} & 41.33 \\
& $M_2$ & \textbf{\underline{28.90}} & 23.45 & 23.40 & 24.83 & 25.14 \\
& $M_3$ & \textbf{\underline{67.94}} & 46.20 & 17.03 & 50.16 & 45.33\\
\bottomrule
\end{tabular}
\end{table}

\subsection{RQ1: Impact on Decision Outcome}
\label{s:rq1}

\begin{figure}[t]
	\centering
	\includegraphics[width=\columnwidth]{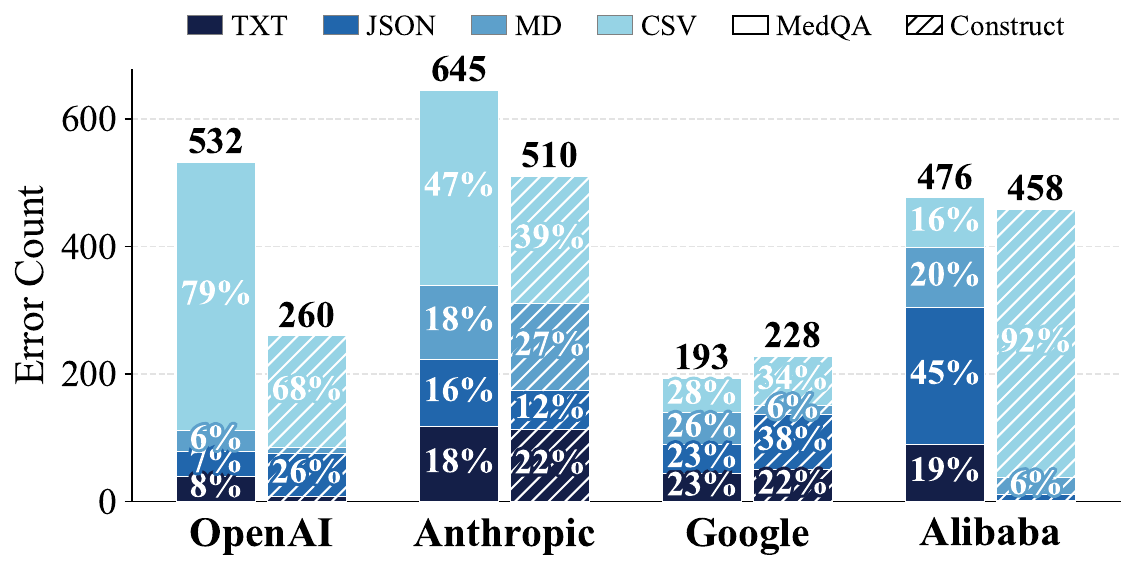}
	\caption{Erroneous Results Across Formats.}
	\label{fig:rq1_flip}
\end{figure}

\begin{figure}[t]
	\centering
	\includegraphics[width=\columnwidth]{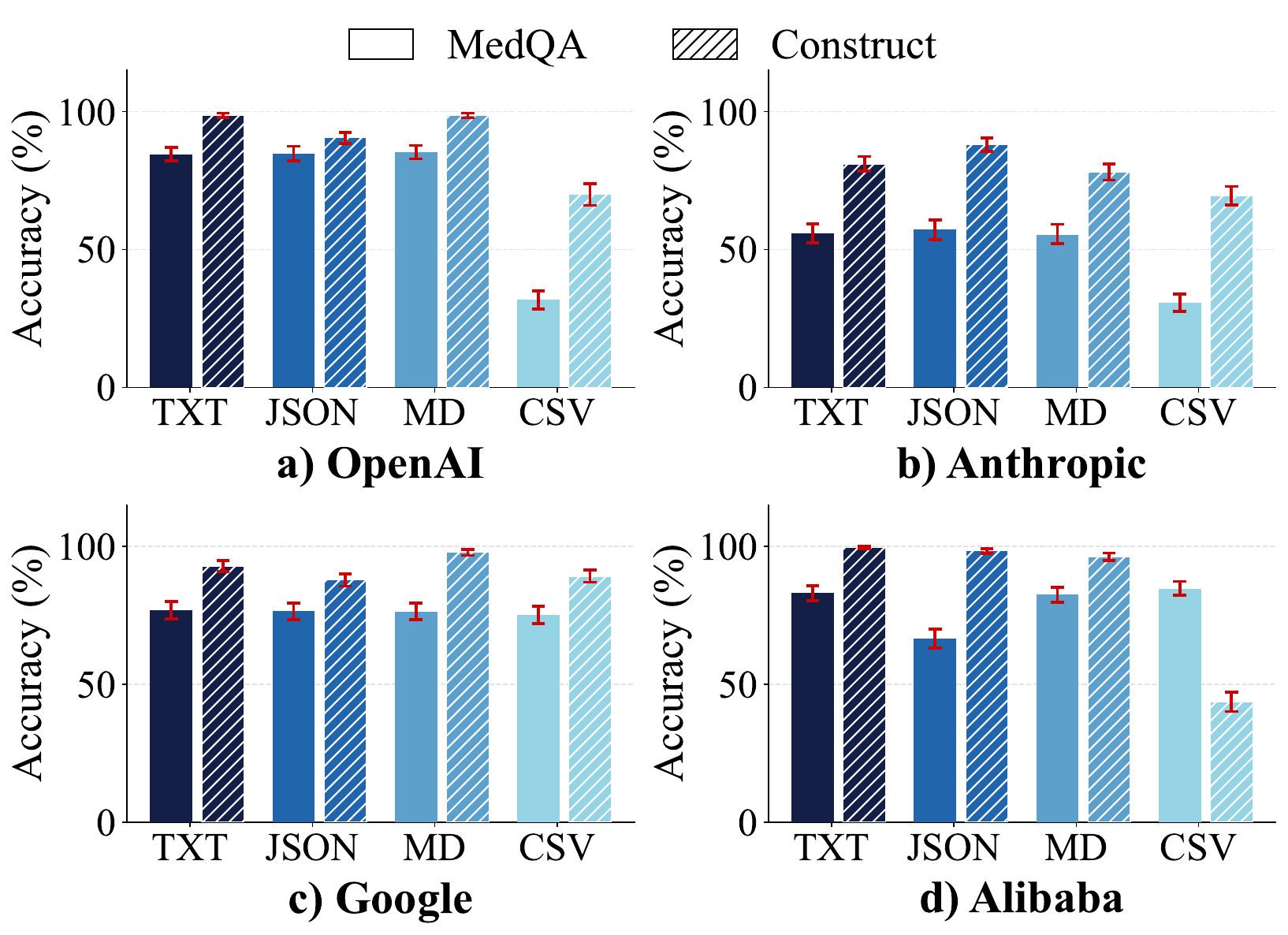}
	\caption{Accuracy Across Formats and Workflows on MedQA and Construct Dataset.}
	\label{fig:rq1_acc}
\end{figure}

To systematically investigate the impact of input formats on the decision outcomes of LLM document workflows, we collect and analyze a total of 12,000 execution response groups across four workflows and four datasets.
Each group contains four workflow responses to the same semantic input in four formats.
We first compute the MR1 violation rate (\(M_1\)) for each workflow and analyze the format distribution of erroneous results among instances that trigger correctness flips, thereby revealing the underlying error patterns of the workflows.
\upd{Response to Reviewer-C-C2}{
\autoref{tab:mrv_provider} presents the MRV results across workflows and datasets, where bold and underlined number in each row marks the workflow with the highest violation rate for each metric.
}
\autoref{fig:rq1_flip} shows the distribution of erroneous results across formats, where the X-axis denotes the workflow provider and bars with and without dashed fills correspond to different datasets.
We then quantitatively evaluate the impact of format variations on the overall task performance (i.e., accuracy) of workflows on the MedQA and Construct datasets, which possess ground-truth answers, to examine how MR1 violations translate into functional degradation.
\autoref{fig:rq1_acc} shows the accuracy of each workflow under different input formats.

\noindent
\(\bullet\)
{\bf Analysis of MR1 Violation.}
\autoref{tab:mrv_provider} reveals that MR1 violations are pervasive across all experimental settings.
Across the experimental datasets and workflows, 41.33\% of instances exhibit inconsistent decision outcomes between at least one pair of formats, confirming that format variations systematically disrupt outcome invariance.
Notably, the Google workflow on Credit Card reaches the highest $M_1$ of 91.60\%, indicating that mere format switching can severely undermine the decision consistency of this workflow.

To verify whether these violations correspond to compromised decision correctness rather than the random wandering among different incorrect options, we delve into the subset of instances where format changes trigger correctness flips (i.e., under the same semantics, the results are correct in some input formats but incorrect in others).
As shown in~\autoref{fig:rq1_flip}, the erroneous results are not uniformly distributed across all formats.
They are noticeably concentrated in structured formats such as \texttt{CSV}.
Specifically, among the 3,302 instances, the \texttt{CSV} format input yields the highest error rate, reaching up to 91.92\% across the four workflows.
% This indicates that a large proportion of the MR1 violation cases result in incorrect decision outputs for input under \texttt{CSV} format.
In contrast, text-based representations like \texttt{TXT} and \texttt{MD} appear more frequently on the correct side, with the lowest error rate being only 0.22\%.
This indicates that the errors induced by MR1 violations exhibit clear directionality, with certain formats (e.g., \texttt{CSV}) being systematically more susceptible to triggering erroneous decisions.
This observation is statistically confirmed by Cochran's Q-tests across all workflow-dataset combinations (all $p < 0.001$), and paired McNemar's tests with Holm correction further verify that the error rates of \texttt{CSV} are significantly higher than those of \texttt{TXT} and \texttt{MD}.

\finding{The violations on MR1 are widespread at the instance level, with an average of 41.33\% on four datasets.
Furthermore, error cases induced by MR1 violations are significantly concentrated in the \texttt{CSV} formats.}

\noindent
\(\bullet\)
{\bf Analysis of Overall Performance.}
We further study how format variations manifest as the degradation of overall task performance.
As illustrated in~\autoref{fig:rq1_acc}, format variations directly alter the workflow's accuracy on the datasets.
Specifically, text-based representations such as \texttt{TXT} and \texttt{MD} typically lead to higher accuracy across different workflows, whereas \texttt{CSV} and certain \texttt{JSON} representations are more prone to triggering significant degradation.
For example, on the Construct dataset, the average accuracy across workflows for \texttt{TXT} is 93.05\%, whereas \texttt{CSV} drops to 67.92\%.
This confirms that in LLM document workflows, even when semantic content and evaluation criteria remain constant, the input document format alone can substantially alter the system's task performance.

More importantly, the degradation pattern is not determined by the format but reflects a significant format-workflow interaction.
The worst-performing format varies across workflows.
On the Construct dataset, the OpenAI workflow achieves its best accuracy with \texttt{MD} yet its worst with \texttt{CSV}, while the Alibaba workflow exhibits the most severe degradation under \texttt{CSV}, with the maximum performance gap reaching 56.00\%.
% This demonstrates that format-induced degradation is closely tied to each workflow's document ingestion pathways, parsing mechanisms, and context organization strategies, rather than being an intrinsic property of any single format.
This demonstrates that format degradation is related not only to document representation but also closely tied to the workflow's native access pathways, parsing methods, and context organization mechanisms.
It is the result of the combined effect of the format and the specific workflow design

Furthermore, we observe that similar accuracy across formats does not imply a low MR1 violation rate.
% For instance, when the Anthropic workflow processes MedQA, the overall accuracy of \texttt{MD} and \texttt{JSON} is exceptionally close (26.90\% and 28.50\%, respectively), yet the MR1 violation rate between them still reaches 9.53\%.
For instance, when the Anthropic workflow processes MedQA, the overall accuracy of \texttt{MD} and \texttt{JSON} is exceptionally close (55.43\% and 57.16\%, respectively), yet the MR1 violation rate between the two formats reaches 16.14\%.
The root cause is that instances answered incorrectly under one format may coincidentally be answered correctly under the other, and vice versa, so that correct and incorrect results cancel each other out in aggregate statistics.
This implies that relying solely on average accuracy can substantially underestimate format-induced risks by masking significant behavioral inconsistency at the instance level, further highlighting the necessity of metamorphic testing for instance-level verification.

\finding{The changes in document formats can cause the accuracy of LLM workflows to drop by up to 56.00\%.
Furthermore, proximity in average accuracy does not guarantee behavioral consistency of the workflow across different formats.
Although some formats exhibit similar overall performance, they may still contain a massive number of result differences concealed by the average values.}

\subsection{RQ2: Impact on Reasoning Evidence}
\label{s:rq2}

\begin{table}[t]
\centering
\caption{Evidence Drifts Under Different Formats.}
\label{tab:rq2_format_evidence}
\scriptsize
\tabcolsep=1.5pt
\begin{tabular}{cccccccccccccc}
\toprule
\textbf{Dataset} & \textbf{Format}
& \multicolumn{3}{c}{\textbf{OpenAI}}
& \multicolumn{3}{c}{\textbf{Anthropic}}
& \multicolumn{3}{c}{\textbf{Google}}
& \multicolumn{3}{c}{\textbf{Alibaba}} \\
\cmidrule(r){3-5}\cmidrule(r){6-8}\cmidrule(r){9-11}\cmidrule(r){12-14}
& & \textbf{JS} & \makecell{\textbf{$\Delta$Acc}\\\textbf{(\%)}} & \makecell{\textbf{Flip}\\\textbf{(\%)}} 
  & \textbf{JS} & \makecell{\textbf{$\Delta$Acc}\\\textbf{(\%)}} & \makecell{\textbf{Flip}\\\textbf{(\%)}} 
  & \textbf{JS} & \makecell{\textbf{$\Delta$Acc}\\\textbf{(\%)}} & \makecell{\textbf{Flip}\\\textbf{(\%)}} 
  & \textbf{JS} & \makecell{\textbf{$\Delta$Acc}\\\textbf{(\%)}} & \makecell{\textbf{Flip}\\\textbf{(\%)}}  \\
\midrule
\multirow{3}{*}{MedQA}
& JSON & 0.18 & 0.27 & 3.33 & 0.18 & 1.47 & 10.71 & 0.14 & -0.13 & 6.72 & 0.15 & -16.56 & 33.24\\
& MD   & 0.10 & 0.80 & 1.87 & 0.13 & -0.41 & 4.54 & 0.10 & -0.81 & 2.43 & 0.14 & -0.53 & 6.68 \\
& CSV  & 0.52 & -52.34 & 65.70 & 0.68 & -25.20 & 74.27 & 0.19 & -1.21 & 14.15 & 0.32 & 1.60 & 20.05 \\
\midrule
\multirow{3}{*}{Construct}
& JSON & 0.15 & -7.88 & 9.21 & 0.19 & 6.94 & 14.42 & 0.16 & -4.95 & 13.50 & 0.17 & -1.20 & 1.20 \\
& MD   & 0.12 & 0.00 & 2.14 & 0.63 & -3.07 & 23.63 & 0.16 & 4.97 & 8.20 & 0.09 & -3.33 & 3.60 \\
& CSV  & 0.32 & -28.76 & 29.78 & 0.21 & -11.63 & 22.86 & 0.19 & -3.75 & 13.92 & 0.33 & -56.00 & 56.00 \\
\bottomrule
\end{tabular}
\end{table}

\begin{figure}[t]
\centering	\includegraphics[width=\columnwidth]{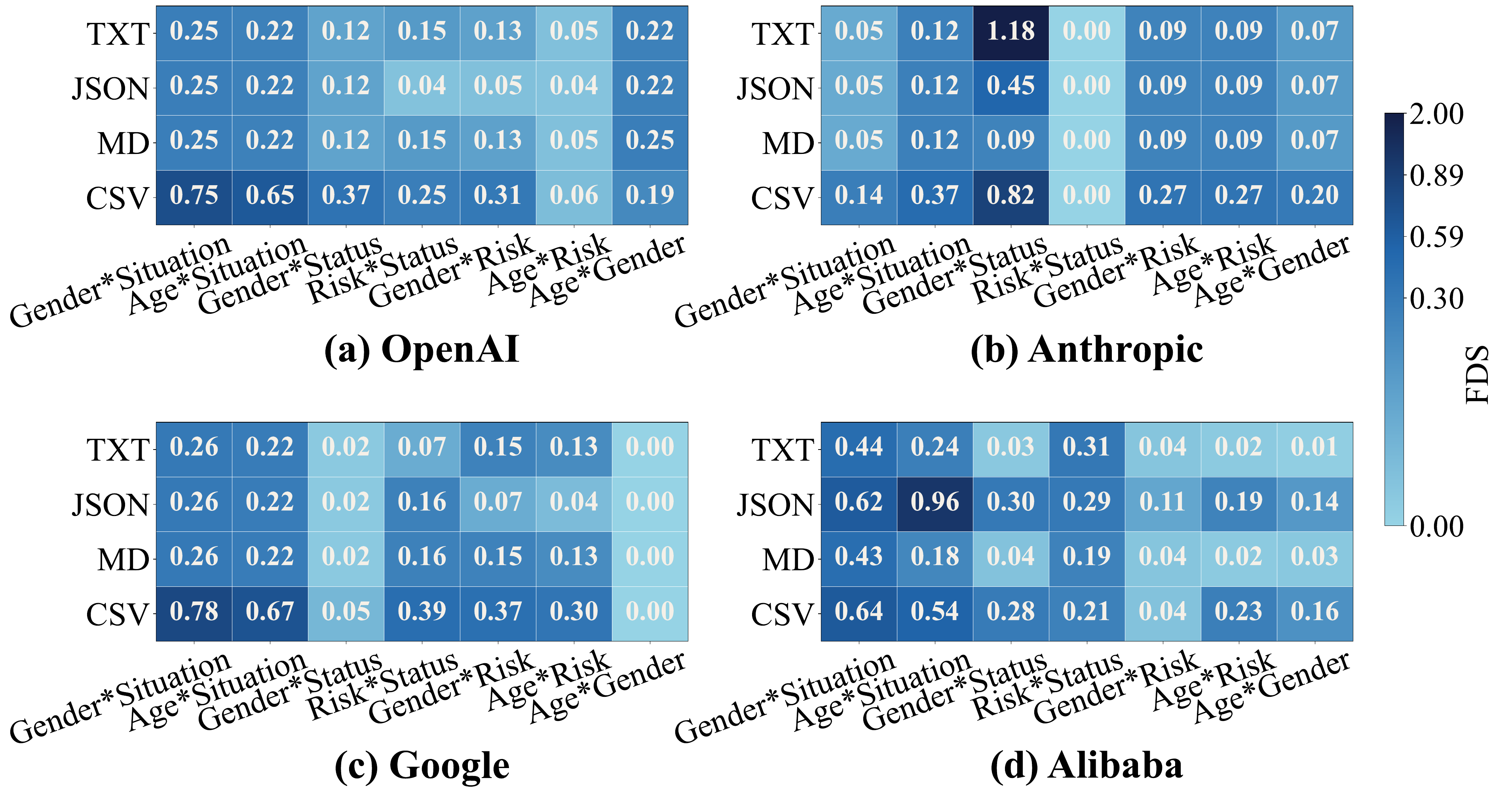}
\caption{Format Disparity Score Heatmap on DiscrimEval Dataset.}
\label{fig:rq2_heatmap}
\end{figure}

To investigate how input formats affect the evidence extraction and reasoning process of LLM workflows, we collect and analyze a total of 12,000 evidence groups, each comprising the workflow's evidence sets obtained under four different input formats.
We first measure the MR2 violation rate (\autoref{tab:mrv_provider}) and analyze the co-occurrence relationship between MR2 and MR1 violations.
Subsequently, using the \texttt{TXT} format, which most closely resembles plain-text prompts, as the baseline, we employ Jensen-Shannon (JS) divergence to quantify the degree of evidence distribution shift under different formats and analyze how this drift correlates with performance degradation.
\autoref{tab:rq2_format_evidence} reports the JS divergence (`JS'), accuracy difference (`$\Delta$Acc'), and decision flip rate (`Flip') across formats and datasets.
Note that a positive `$\Delta$Acc' indicates that the given format outperforms the \texttt{TXT} baseline, and `Flip' counts the proportion of decision output in the given format that is different from the results of \texttt{TXT} baseline.
Finally, we conduct a paired-factor sensitivity analysis to examine format-induced fairness risks.
\autoref{fig:rq2_heatmap} presents heatmaps of FDS on the DiscrimEval dataset, where the X-axis indicates factor pairs.
% and the Y-axis shows formats.

\noindent
\(\bullet\)
{\bf Analysis of MR2 Violation.}
\autoref{tab:mrv_provider} reveals that evidence drift is a systemic phenomenon within end-to-end LLM document workflows.
On average, 25.14\% of instances trigger an MR2 violation across all settings.
The severity varies considerably across workflows and datasets.
% For example, the Google workflow on DiscrimEval reaches the highest $M_2$ of 53.10\%, while the Alibaba workflow on Construct exhibits only 1.20\%.
\upd{Response to Reviewer-B-C5}{Cochran's Q-tests confirm that inter-workflow differences in $M_2$ are statistically significant across all four datasets (all $p < 0.001$)~\cite{cochran1950the,McNemar1947note,sture1979a}.}
%  and Construct ($Q=257.18$)
To reveal the relationship between evidence drift and decision changes, we conduct a co-occurrence analysis of MR1 and MR2 violations.
Across all four datasets, 63.56\% of MR2-violating instances simultaneously violate MR1, confirming that evidence drift is strongly associated with decision changes.
However, 36.44\% of MR2 violations are \textit{silent violations}, where the workflow arrives at the same final decision but cites drastically different evidence under different formats.
\upd{Response to Reviewer-B-C5}{The silent violation rate varies substantially across datasets, from 11.11\% on MedQA to 55.33\% on DiscrimEval.}
Such silent violations are particularly concerning in high-stakes scenarios, as they undermine the interpretability and auditability of system decisions without any observable change in the output.
To further characterize the distributional features of these violations, we use the \texttt{TXT} format as a baseline and conduct pairwise analysis across formats.
We observe that 52.21\% and 24.35\% of evidence changes occur under the \texttt{CSV} and \texttt{JSON} formats, suggesting that structured formats are more prone to redistributing the salience of fields during document ingestion, thereby altering the core evidence relied upon by the workflow.

\finding{The average MR2 violation rate reaches 25.14\%, indicating that format changes frequently induce workflows to extract different information cues for reasoning.
Even worse, 36.44\% of these violations are \textit{silent}, whose reasoning logic dislocations are masked by consistent decision outcomes, posing direct threats to the interpretability and trustworthiness of LLM document workflows.}

% Dataset     | Total MR2 | Both(MR1&MR2) | MR2-only(silent) | MR1-only
% DiscrimEval |   291     |   130         |   161            |  170
% Construct   |   199     |   115         |    84            |  300
% Credit Card |   286     |   122         |    30            |  332
% MedQA       |   162     |   144         |    18            |  745

\noindent
\(\bullet\)
{\bf Analysis of Evidence Drift.}
The results in~\autoref{tab:rq2_format_evidence} reveal that evidence drift (high `JS') is systematically associated with higher decision flip rates (high `Flip') and lower relative accuracy (low `$\Delta$Acc', where a higher value indicates better performance relative to \texttt{TXT}).
\upd{Response to Reviewer-B-C5}{Spearman correlation analysis confirms this pattern across all settings ($\rho(\mathrm{JS}, \Delta\mathrm{Acc})=-0.470$, $p < 0.05$), with the effect strengthening on the MedQA datasets (up to $\rho=-0.683$).
In the most severe case, the Anthropic workflow under \texttt{CSV} on MedQA reaches a flip rate of 74.27\%.
Notably, the worst-case format varies across workflows.
On MedQA, all four workflows exhibit the most severe drift under \texttt{CSV}, whereas on Construct, the worst-case format differs by workflow (e.g., \texttt{MD} for Anthropic, \texttt{CSV} for OpenAI and Alibaba).}
This confirms that evidence drift is not attributable to a single format but arises from the interaction between document formats and the specific processing mechanisms of each workflow.

\begin{figure}[t]
	\centering
	\includegraphics[width=\columnwidth]{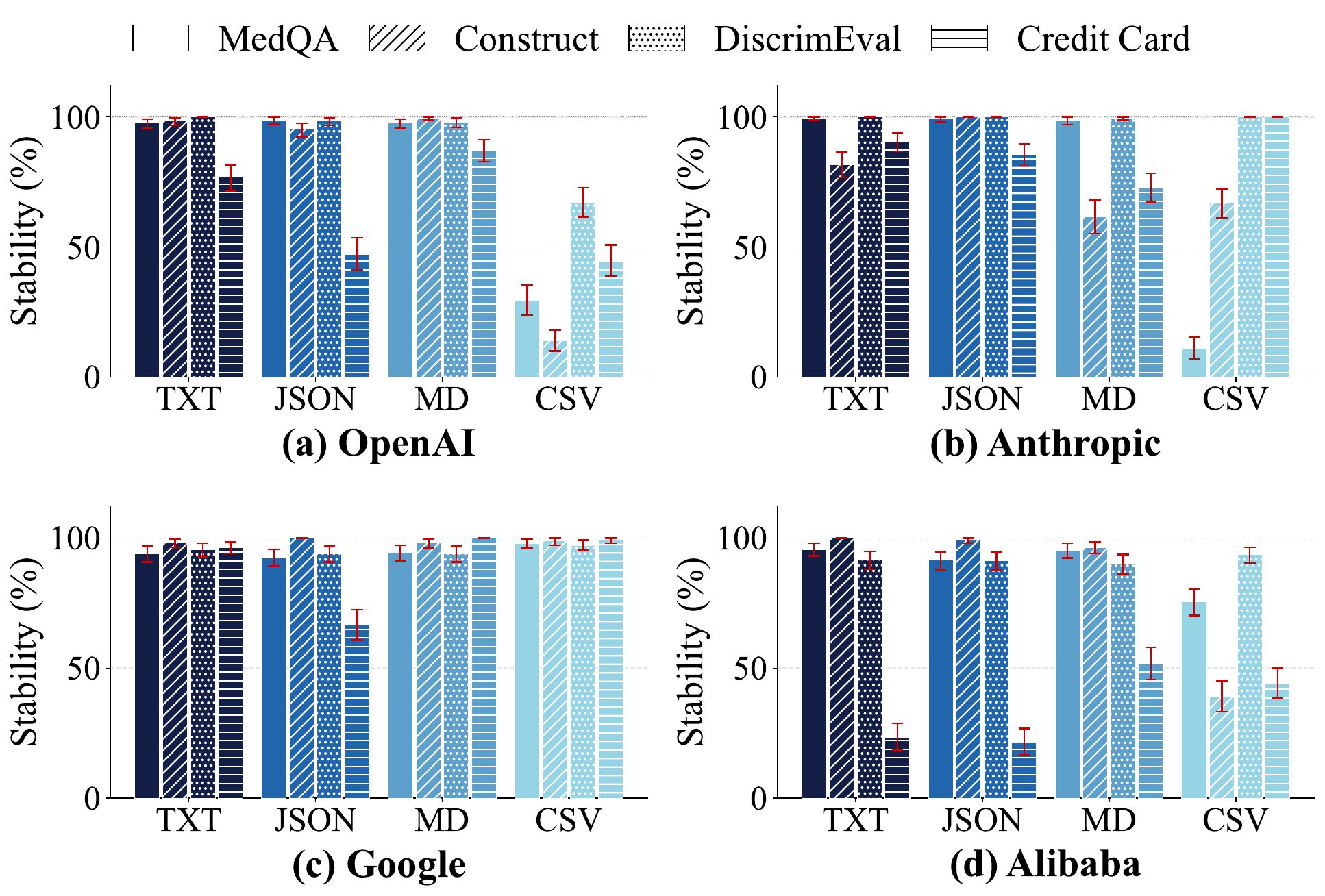}
	\caption{Stability ($Stb$) Across Different Datasets and Workflows.}
	\label{fig:rq3}
\end{figure}
%\todo{xianyun, this should be a barchart like fig:rq1acc.}

To further investigate the fairness risks of evidence drift, we conduct a paired-factor sensitivity analysis across all four datasets and four workflows.
By fixing the decision question and varying sensitive attributes (e.g., gender and age), we compute the FDS to quantify the maximum group-level decision disparity induced by each format.
Overall, we observe that format-induced fairness risks are pervasive.
Elevated FDS values appear across all four datasets, with structured formats (particularly \texttt{CSV}) consistently triggering the most severe group disparities.
Due to space limitations, \autoref{fig:rq2_heatmap} illustrates part of the heatmap example on the DiscrimEval dataset, where `Gender$\times$Status' achieves the largest FDS on \texttt{TXT} format.
Complete heatmaps for all datasets and workflows are released~\cite{our_repo}.
\upd{Response to Reviewer-B-C5}{On the Construct dataset, three workflows reach their peak FDS under \texttt{CSV} on the `Age$\times$Job' pair (FDS up to 2.00), whereas other formats produce identical decision rates across these groups, demonstrating that a single format change to \texttt{CSV} can introduce substantial group disparity that is entirely absent under other formats. Similar patterns appear on the Credit Card and MedQA datasets, where specific formats disproportionately amplify disparity on sensitive attribute pairs such as `Gender$\times$Marital' and gender-related factors. These results indicate that when a workflow parses specific syntactic structures, its reasoning may attend to non-task-relevant sensitive attributes, ultimately inducing systemic decision biases.}

\upd{Response to Reviewer-B-Q2}{
To further illustrate how evidence drift occurs, we study a representative case on the Credit Card dataset.
For the exact same record (i.e., `a 50-year-old married male whose account was inactive for five months with only one late payment across six months'), the Google workflow yields opposite decisions depending on the format. 
Under \texttt{MD}, the workflow returns `YES' (80\% confidence) and predicts a high risk of default for this case by listing business-relevant evidence (e.g., the payment delay and the revolving balance).
Conversely, under \texttt{JSON}, the workflow flips to `NO' (95\% confidence), and the evidence shifts to demographic attributes, such as age and gender (`50-year-old male') and marital status (`married').
The workflow finally interprets this record as `only one late payment' and concludes with low default risk.
Both the motivation case (\autoref{fig:moti}) and the above case illustrate that format-induced evidence drift can redirect the workflow's attention from task-relevant financial signals to non-task-relevant demographic attributes, silently introducing fairness violations into high-stakes credit decisions.
}
% Together with the aggregate JS and FDS results above, this example explains how format-induced changes in evidence selection can introduce fairness violations in high-risk scenarios.}
%  at the group level

\finding{Format-induced evidence drift is significantly correlated with downstream decision flips and performance degradation.
Furthermore, specific structured formats can perturb the salience of factual cues, amplifying decision disparities across demographic groups (FDS up to 2.00) and silently introducing fairness violations in high-stakes scenarios.}

  % Max FDS by dataset/model:
  %   construct/Qwen:   FDS=2.00  csv   age*job（|0.333-0.083|/0.083=3.0→clip=2.00为了和其他热力图对齐选择了用2.00作为可视化上限）
  %   construct/GPT:    FDS=1.667  csv   age*job
  %   construct/Gemini: FDS=1.667  csv   age*job
  %   construct/Claude: FDS=0.487  md    gender*marriage
  %   financial/Gemini: FDS=1.276  txt   Sex*Marital
  %   financial/Qwen:   FDS=0.714  txt   Sex*Education
  %   financial/GPT:    FDS=0.612  json  Sex*AgeBin
  %   financial/Claude: FDS=0.637  csv   Sex*AgeBin
  %   medqa/GPT:        FDS=0.534  csv   Gender*HasDiagnostic
  %   medqa/Claude:     FDS=0.818  txt   HasDiagnostic*HasPhysicalExam
  %   medqa/Gemini:     FDS=0.516  csv   Gender*HasPhysicalExam
  %   medqa/Qwen:       FDS=0.731  md    Gender*HasPhysicalExam

\subsection{RQ3: Impact on Execution Stability}
\label{s:rq3}

%\todo{xianyun, carefully check the numbers, statistical results, and findings, and update the 'xxx' placeholders in this section based on the latest experimental data.}

To investigate how input format affects the execution stability of LLM document workflows, we collect and analyze 3,000 groups ($k=3$) of repeated execution results across four workflows and four datasets.
We first measure the MR3 violation rate to quantify the extent to which format variation undermines decision stability.
\autoref{tab:mrv_provider} reports $M_3$ for each workflow across different datasets.
We then compare the stability metric $Stb$ (\autoref{s:setup}) across formats at the dataset level to discuss the practical deployment risks.
\autoref{fig:rq3} presents the distribution of $Stb$ for each workflow across different formats and datasets.

\noindent
\(\bullet\)
{\bf Analysis of MR3 Violation.}
The results in~\autoref{tab:mrv_provider} indicate that MR3 violations are prevalent across workflows.
On average, 45.33\% of instances trigger an MR3 violation, demonstrating that switching input formats not only alters single-execution decision outcomes (\autoref{s:rq1}) but also systematically undermines the determinism of the reasoning process within workflows.
Moreover, MR3 violations are highly unevenly distributed across workflows.
For example, on Construct, the OpenAI workflow reaches an $M_3$ of 86.40\%, whereas the Google workflow exhibits only 4.80\%.
Cochran's Q-tests on aligned instances confirm that these inter-workflow differences are statistically significant across all four datasets (all $Q > 338$, $p < 0.001$).
This substantial disparity suggests that the degree to which format variation disrupts execution stability is highly dependent on each workflow's underlying mechanisms, including file parsing paths and context organization strategies, whose implementation-level differences are further amplified across repeated executions.

\finding{The average MR3 violation rate reaches 45.33\%, indicating that input format variation can systematically alter the execution consistency of instances. Furthermore, the violations exhibit a significantly uneven distribution across workflows, with $M_3$ reaching 67.94\% on the OpenAI workflow and 17.03\% on the Google workflow.}

\noindent
\(\bullet\)
{\bf Analysis of Stability Degradation.}
As shown in~\autoref{fig:rq3}, stability degradation does not occur uniformly across all formats but instead exhibits a pronounced clustering effect within specific formats.
\upd{Response to Reviewer-B-C5}{Taking the OpenAI workflow on Construct as an example, the system maintains high stability under \texttt{TXT}, \texttt{JSON}, and \texttt{MD} (all $Stb > 95\%$), but $Stb$ drops sharply to 14.00\% under \texttt{CSV}.}
Anthropic pipeline achieves the lowest $Stb$ of 11.00\% on the \texttt{CSV} format.
In contrast, the Google and Alibaba workflows maintain relatively higher stability.
This comparison suggests that format-induced stability degradation is jointly driven by the interaction between specific document representations and the workflow's underlying processing mechanisms.

This finding carries significant implications for LLM-driven software engineering practices.
Under specific structured formats, the underlying processes of the workflow (e.g., file parsing and serialization) propagate syntactic differences in input representations to the model's reasoning layer, systematically amplifying the inherent stochasticity of inference.
In high-stakes scenarios such as automated medical triage and financial statement auditing~\cite{lawrence2024opportunities,xinyi2024large}, such unpredictable fluctuations will severely compromise the reproducibility of software systems and substantially increase the costs of debugging and verification, posing a fundamental challenge to the trustworthy deployment of LLM document workflows in production environments.

\finding{Document format is a critical variable in determining the execution stability of LLM workflows. Specific structured formats (e.g., \texttt{CSV}) can amplify the stochasticity during model reasoning, causing $Stb$ to drop to as low as 11.00\%, thereby posing a severe reliability threat to the production deployment of document workflows in high-stakes scenarios.}

\subsection{RQ4: Mitigation Strategies}
\label{s:rq4}

\begin{table}[t]
\centering
\caption{\upd{Response to Reviewer-B-C5}{Effect of Mitigation Strategies on MR Violation Rates \scriptsize{(Bold and underlined marks the better result between two strategies)}.}}
\label{tab:rq4}
\scriptsize
\tabcolsep=1.5pt
\begin{tabular}{cc cc cc cc cc cc}
\toprule
\textbf{Dataset} & $\Delta$\textbf{MRV} & \multicolumn{2}{c}{\textbf{OpenAI}} & \multicolumn{2}{c}{\textbf{Anthropic}} & \multicolumn{2}{c}{\textbf{Google}} & \multicolumn{2}{c}{\textbf{Alibaba}} & \multicolumn{2}{c}{\textbf{Avg.}} \\
\cmidrule(r){3-4} \cmidrule(r){5-6} \cmidrule(r){7-8} \cmidrule(r){9-10} \cmidrule(r){11-12}
& & Vot. & Rte. & Vot. & Rte. & Vot. & Rte. & Vot. & Rte. & Vot. & Rte. \\
\midrule
\multirow{3}{*}{MedQA}
& $\Delta M_1$ & 2.80 & \textbf{\underline{65.20}} & 21.71 & \textbf{\underline{76.67}} & 4.93 & \textbf{\underline{17.38}} & -4.27 & \textbf{\underline{41.87}} & 8.43 & \textbf{\underline{50.28}}  \\
& $\Delta M_2$  & 0.00 & \textbf{\underline{16.80}} & 0.00 & \textbf{\underline{12.00}} & 0.00 & \textbf{\underline{19.70}} & 0.00 & \textbf{\underline{17.20}} & 0.00 & \textbf{\underline{16.43}} \\
& $\Delta M_3$ & \textbf{\underline{71.37}} & 68.95 & \textbf{\underline{88.80}} & 87.97 & \textbf{\underline{15.73}} & 9.68 & \textbf{\underline{31.85}} & 7.25 & \textbf{\underline{51.94}} & 43.46 \\
\midrule
\multirow{3}{*}{Construct}
& $\Delta M_1$ & 13.73 & \textbf{\underline{29.33}} & 10.93 & \textbf{\underline{44.53}} & -0.13 & \textbf{\underline{21.47}} & -1.60 & \textbf{\underline{57.20}} & 5.73 & \textbf{\underline{38.13}} \\
& $\Delta M_2$ & 0.00 & \textbf{\underline{7.60}} & 0.00 & \textbf{\underline{64.90}} & 0.00 & \textbf{\underline{6.40}} & 0.00 & \textbf{\underline{1.20}} & 0.00 & \textbf{\underline{20.03}} \\
& $\Delta M_3$ & \textbf{\underline{86.40}} & 86.00 & \textbf{\underline{64.80}} & \textbf{\underline{64.80}} & \textbf{\underline{4.80}} & 2.80 & \textbf{\underline{63.20}} & \textbf{\underline{63.20}} & \textbf{\underline{54.80}} & 54.20 \\
\midrule
\multirow{3}{*}{Avg.}
& $\Delta M_1$ & 8.27 & \textbf{\underline{47.27}} & 16.32 & \textbf{\underline{60.60}} & 2.40 & \textbf{\underline{19.42}} & 1.33 & \textbf{\underline{49.53}} & 7.08 & \textbf{\underline{44.21}} \\
& $\Delta M_2$ & 0.00 & \textbf{\underline{12.20}} & 0.00 & \textbf{\underline{38.45}} & 0.00 & \textbf{\underline{13.05}} & 0.00 & \textbf{\underline{9.20}} & 0.00 & \textbf{\underline{18.23}} \\
& $\Delta M_3$ & \textbf{\underline{78.89}} & 77.48 & \textbf{\underline{76.80}} & 76.39 & \textbf{\underline{10.27}} & 6.24 & \textbf{\underline{47.53}} & 35.23 & \textbf{\underline{53.37}} & 48.83 \\
\bottomrule
\end{tabular}
\end{table}

% \begin{table}[t]
% 		\caption{Baseline diagnostics on empirically worst formats.}\label{tab:baseline_worst}
% 		\label{tab:baseline_diag}
% 		\centering
% 		\scriptsize
% 		\begin{adjustbox}{max width=\columnwidth}
% 			\begin{tabular}{@{}l l l c c c c c c@{}}
% 				\toprule
% 				& \textbf{Model} & \textbf{WorstFmt} & $\mathrm{Acc}_{rep1}$ & $\mathrm{Std}$ & $\mathrm{Acc}_{vote}$ & \textbf{VoteGain} & \textbf{AgreeRate} & $P_{\mathrm{all\ wrong}}$ \\
% 				\midrule
% 				\multirow{4}{*}{\vhead{Construct}} & Claude & XLSX & 0.436 & 0.042 & 0.508 & 0.072 & 0.468 & 0.468 \\
% 				& GPT & CSV & 0.492 & 0.259 & 0.568 & 0.076 & 0.436 & 0.023 \\
% 				& Gemini & XLSX & 0.560 & 0.008 & 0.572 & 0.012 & 0.984 & 0.990 \\
% 				& Qwen & XLSX & 0.180 & 0.007 & 0.176 & -0.004 & 0.864 & 0.907 \\
% 				\midrule
% 				\multirow{4}{*}{\vhead{MedQA}} & Claude & CSV & 0.368 & 0.069 & 0.264 & -0.104 & 0.116 & 0.639 \\
% 				& GPT & CSV & 0.312 & 0.024 & 0.276 & -0.036 & 0.332 & 0.680 \\
% 				& Gemini & XLSX & 0.204 & 0.007 & 0.212 & 0.008 & 0.832 & 0.987 \\
% 				& Qwen & JSON & 0.664 & 0.002 & 0.664 & 0.000 & 0.912 & 0.917 \\
% 				\bottomrule
% 			\end{tabular}
% 		\end{adjustbox}
% 	\end{table}

To mitigate format-induced errors in LLM workflows, we design two lightweight training-free strategies from the user's perspective.
\textcircled{1} \textbf{Vote Aggregation} is a generic strategy inspired by prior work~\cite{wang2023selfconsistency}. It keeps the input format unchanged and aggregates decisions from multiple executions (i.e., three in our experiments) via majority voting, aiming to reduce random errors through posterior statistics.
\upd{Response to Reviewer-C-Q1}{
\textcircled{2} \textbf{Format Routing} is a targeted strategy designed specifically for format-induced errors.
It inserts a lightweight routing layer upstream of the workflow that converts every input document to the empirically optimal format before inference.
Concretely, this layer maps the input task domain to the optimal format for the current workflow based on offline empirical results. For example, a \texttt{CSV} MedQA input to the OpenAI workflow is automatically converted to \texttt{MD} (its optimal format) prior to inference.
Leveraging the pre-collected empirical results rather than per-document LLM analysis, this approach introduces negligible overhead.
Detailed implementations for both strategies are in~\cite{our_repo}.}
We apply both strategies to four workflows on Construct and MedQA, analyzing 8,000 sets of instances.
\upd{Response to Reviewer-B-C5}{\autoref{tab:rq4} summarizes the results, where $\Delta$MRV denotes the reduction in violation rate after applying each strategy, and `Vot.' and `Rte.' abbreviate the vote aggregation and format routing strategies, respectively.
Bold and underlined highlight the results of the better of the two strategies.}

\noindent
\(\bullet\)
{\bf Analysis of Mitigation Methods.}
As shown in~\autoref{tab:rq4}, the vote aggregation strategy provides only limited improvement on decision outcomes.
On Construct, the average $\Delta M_1$ achieved by voting is merely 5.73\%.
On the MedQA dataset, the voting strategy even degrades the Alibaba workflow ($\Delta M_1 = -4.27\%$), confirming that when format-induced bias is embedded in the input representation, repeated sampling merely reproduces inferior reasoning paths rather than generating corrective information~\cite{wang2023selfconsistency,dhuliawala2024chain}.
Notably, $\Delta M_2$ remains 0.00\% across all voting configurations because voting operates within a single format, aggregating decisions from repeated executions without altering evidence cues, thus leaving $M_2$ violations entirely unaffected.

In contrast, format routing achieves substantial and consistent mitigation across all three MRs, with average reductions of 44.21\%, 18.23\%, and 48.83\% on $M_1$, $M_2$, and $M_3$, respectively, outperforming the voting strategy by 37.13\% on $M_1$.
For example, on Construct, routing raises the accuracy of the OpenAI workflow from 69.93\% under \texttt{CSV} to 98.67\% under \texttt{MD} (gain = 28.74\%, 95\% CI: [22.00\%, 36.00\%]).
% The reduction in $M_2$ (18.23\%) further indicates that routing to the empirically optimal format not only corrects decision outcomes but also stabilizes the evidence cues surfaced by the workflow, reducing cross-format evidence divergence.
% This contrast is consistent with our findings in~\autoref{s:rq2} and~\autoref{s:rq3}.
\upd{Response to Reviewer-C-Q1}{
This contrast in effectiveness, which aligns with our findings in~\autoref{s:rq2} and~\autoref{s:rq3}, stems from the fundamental difference in intervention timing.
Format-induced errors originate directly in the document processing mechanisms of workflows, where different formats produce structurally divergent context representations that inject noise into the reasoning layer.
Format routing addresses this root cause proactively by converting documents into an empirically stable format before processing.
As evidenced by the 18.23\% reduction in $M_2$, this representation-level intervention not only corrects macroscopic decision outcomes but also stabilizes the underlying evidence cues.
Conversely, majority voting merely performs post-hoc aggregation.
If a specific format systematically degrades the reasoning process, repeated executions will only reproduce those identical errors without alleviating the inherent representation-level bias.
}

\finding{The format routing strategy effectively mitigates format-induced errors across all three MRs, achieving average reductions of 44.21\%, 18.23\%, and 48.83\% in $M_1$, $M_2$, and $M_3$.
This confirms that format-induced errors originate in the document processing mechanisms, and intervening at the representation level yields substantial mitigation for decision outcomes, evidence consistency, and execution stability.}

%!TEX root = ../main.tex
\section{Discussion}
\label{sec:discuss}

% \noindent
% \(\bullet\)
% {\bf Broader Significance.}
% \upd{Response to G1-Significance}{
% The findings of this paper extend beyond the four decision-making tasks evaluated in the main study. To assess whether format-induced inconsistency also manifests in open-ended generation tasks, we apply the framework to 15 instances from the CNN/DailyMail summarization dataset~\cite{see2017get}, adapting MR1 to measure semantic similarity between generated summaries and MR2 to track sentence-level source coverage. The results confirm that format variation induces measurable inconsistency in summarization outputs as well, suggesting that format robustness is a broader concern for LLM document workflows beyond high-stakes decision-making domains.
% }

\noindent
\(\bullet\)
{\bf Threats to Validity.}
The generalizability of our findings may be threatened by the following aspects.
\textcircled{1} The experiments are conducted on question-answering and decision-making tasks in high-stakes domains~\cite{singhal2025toward,chu2025domaino1s}, and may not directly capture format-induced degradation in open-ended generation tasks where output correctness is less discretely defined.
However, the underlying principle of our MRs (i.e., semantically equivalent inputs should yield consistent outputs), remains applicable to such tasks through appropriate adaptations.
%  (e.g., using cosine similarity to evaluate the semantic consistency between two generated paragraphs to determine whether the workflow output has substantially changed).
\upd{Response to Metareview-Q3 and Reviewer-A-Q3}{
\textcircled{2} Variations in format expressiveness could theoretically introduce semantic shifts during data generation, confounding the format-induced errors.
Note that our evaluation setting is motivated by real-world practices, where users routinely use diverse formats (e.g., \texttt{CSV} and \texttt{JSON}). 
We mitigate this threat by adhering to real-world usage practices and applying the three transformation constraints outlined in~\autoref{s:data_generation}.}
\upd{Response to Metareview-Q3 and Reviewer-A-Q4}{
Specifically, our framework renders variants from a unified intermediate representation \(C_x\) to isolate task semantics from external formats.    
Furthermore, we invite two independent experts (with at least five years of experience in the SE field) to manually check 20 randomly sampled transformation groups and confirm 100\% semantic consistency across formats. These controls mitigate the threat that MR violations are caused by information loss.
}
\textcircled{3} Our experiments cover four mainstream workflows and four commonly used formats, which may not fully represent some emerging systems, such as skill-based agent systems~\cite{jiang2026sok} or additional document formats.
Note that our testing framework is designed to be extensible. Incorporating new workflows or formats requires only adding the corresponding document conversion operators and API adapters, without modifying the core testing logic.
\upd{Response to Reviewer-B-C4}{\textcircled{4} Executing each configuration only \(k=3\) times may provide a limited view.
To assess this threat, we conduct a supplementary experiment with \(k=5\) on the MedQA and Construct datasets.
While the absolute MR values exhibited marginal shifts, the relative rankings of the most- and least-affected workflows remained identical across both datasets (e.g., on the MedQA dataset, Anthropic remains the highest on MR1/MR3 and Google stays the lowest), confirming that the qualitative observations are stable under a larger \(k\).}
To mitigate these threats and facilitate reproducibility, we release all necessary code, datasets, and results in our repository~\cite{our_repo}.

\noindent
\(\bullet\)
{\bf Future Work.}
\textcircled{1} \textit{Extending MRs and task coverage.}
Current MRs target discrete decision tasks where output correctness can be unambiguously compared.
\upd{Response to Reviewer-A-Q1}{
We have conducted preliminary validation and applied the framework to 15 instances from the CNN-DailyMail summarization dataset~\cite{see2017get}, adapting MR1 to measure semantic similarity between generated summaries and MR2 to track sentence-level source coverage.
We have observed that format-induced inconsistency also persists in common summarization tasks (average MRV of MR1/MR2 across workflows is 48.33\% and 26.67\%).
This further illustrates the significance of this study: this format robustness problem may be widespread in diverse application scenarios, posing risks to the regular use of LLM document workflows.
Future work can systematically design new MRs tailored to generation and summarization tasks using semantic similarity measures, and expand the evaluated workflows to include diverse new pipelines and additional document formats.}
% (e.g., multi-agent systems)
\textcircled{2} \textit{Extending to other workflows.}
\upd{Response to Reviewer-B-Q1}{
While this study deliberately targets production-level workflows to reveal actionable, real-world vulnerabilities regarding format robustness, our framework is extensible to more complex systems.
% Note that this paper aims to evaluate format robustness and reveal format-induced errors in real-world LLM document workflows.
Since our MRs depend on functional properties (e.g., observable final outcomes) rather than architectural complexity, they can be readily adapted to advanced workflows.
However, applying this framework to workflows such as retrieval-augmented generation (RAG) or multi-agent systems introduces new challenges in evidence extraction and tracing.
For instance, quantifying how format variations skew dynamic retrieval distributions in RAG pipelines, or attributing format-induced errors across inter-agent communications, requires advanced tracing methods.
Consequently, constructing customized, modular pipelines to isolate the exact contributions of individual processing stages (e.g., parsing, chunking, and serialization) and conducting systematic root-cause analysis constitutes a critical direction for future work.}
\textcircled{3} \textit{Adaptive format routing.}
The format routing strategy in~\autoref{s:rq4} relies on offline calibration data, and the optimal format may shift as workflows are updated, models are iterated, or task domains change.
Future work can explore adaptive routing mechanisms that dynamically select the optimal input format, e.g., by evaluating workflow performance on a small set of probe samples or training classifiers that adapt to the current workflow state at runtime, thereby reducing reliance on static calibration data and enhancing deployment flexibility.
\textcircled{4} \textit{Scaling repeated-execution evaluation.}
Future work can scale repeated-execution evaluation to more datasets and workflows to obtain finer-grained estimates of stochastic workflow instability.

\section{Conclusion}

This paper presents a format-aware metamorphic testing framework for end-to-end LLM document workflows and conducts a large-scale empirical study across four mainstream workflows and four document formats.
The findings reveal that input format variation constitutes a systematic and previously underexamined threat to the reliability of LLM document workflows.
Merely switching the format of semantically identical content can cause widespread decision flips, severe evidence drift, and significant stability degradation.
Building on the findings, this paper evaluates two lightweight mitigation strategies and shows that the format routing strategy can substantially mitigate the format-induced errors in the workflow.
% We call on that document format should be treated as an important evaluation variable in LLM system testing, and that format-aware safeguards are essential for the trustworthy deployment of LLM document workflows in high-stakes scenarios.
We call on the SE community to treat document format as an important evaluation variable in LLM system testing, and to incorporate format-aware safeguards as a standard practice for the trustworthy deployment of LLM document workflows in high-stakes scenarios.

\newpage

\section{Data Availability Statement}
\label{sec:data}
To follow the Open Science Policy and support reproducibility, we have released the necessary data at~\cite{our_repo}.

\begin{acks}
This research is supported by the National Natural Science Foundation of China (62441237).
This work was also supported by the National Research Foundation, Singapore, and the Cyber Security Agency under its National Cybersecurity R\&D Programme (NCRP25-P04-TAICeN).
This research is also part of the IN-CYPHER Programme and is supported by the National Research Foundation, Prime Minister's Office, Singapore, underits Campus for Research Excellence and Technological Enterprise (CREATE) Programme.
%  as well as the Ministry of Defence (MINDEF), Singapore, through the Defence Science and Technology Agency (DSTA) under Project Agreement No.\ 9025202776.
Any opinions, findings, conclusions, or recommendations expressed in this material are those of the author(s) and do not reflect the views of the National Natural Science Foundation of China, National Research Foundation, or Cyber Security Agency.
\end{acks}

{ \scriptsize
\bibliographystyle{ACM-Reference-Format}
\bibliography{software}
}

% %%
% %% If your work has an appendix, this is the place to put it.
% \appendix

\end{document}